\documentclass[twocolumn,secnumarabic,amssymb,nobibnotes,aps,pre,superscriptaddress]{revtex4-2}

\usepackage{amsmath}
\usepackage{amsfonts}
\usepackage{amssymb}
\usepackage{graphicx}
\usepackage{hyperref}
\usepackage{longtable}
\usepackage{dcolumn}
\usepackage{bm}
\usepackage{epstopdf}
\usepackage{datetime}
\longdate
\usepackage{color}

\newcommand\blue{\textcolor{blue}}

\begin{document}

\title{Membrane tubes with active pumping: water transport, vacuole formation and osmoregulation}

\author{Sami C.\ Al-Izzi}
\email{s.al-izzi@unsw.edu.au}
\affiliation{School of Physics \& ARC Centre of Excellence for the Mathematical Analysis of Cellular Systems, UNSW, Sydney, NSW 2052, Australia.}

\author{Matthew S.\ Turner}
\email{m.s.turner@warwick.ac.uk}
\affiliation{Department of Physics \& Centre for Complexity Science, University of Warwick, Coventry CV4 7AL, UK}

\author{Pierre Sens}
\email{pierre.sens@curie.fr}
\affiliation{Institut Curie, PSL Research University, CNRS, UMR168 Physics of Cells and Cancer, F-75005, Paris, France}

\begin{abstract}
The need for organisms to regulate their volume and osmolarity when surrounded by freshwater is a basic physical challenge for many bacteria, protists and algae. Taking inspiration from the contractile vacuole complex found in many protists, we discuss how simple models of active membrane tubes can give insights into the fluid and active ionic transport properties of such systems. We show that a simple membrane tube with unidirectional ion pumps, and passive ion and water channels, forms a large vacuole due to osmotically-driven water flow and that this can be used to actively pump water out of the cell interior. We discuss the use of this system as a possible minimal method for osmoregulation.
\end{abstract}

\maketitle

%%%%%%%%%%%%
\section{Introduction}
The flux of water across lipid membranes due to osmotic pressure differences is a process vital to the function of many biological systems, from the organism, tissue and organoid level \cite{dasgupta_physics_2018,keener_mathematical_2009,dumortier_hydraulic_2019,verge_physics_2020,cheddadi_coupling_2019,palk_dynamic_2010,torres-sanchez_tissue_2021} down to the scale of a single cell \cite{pilizota_plasmolysis_2013,buda_dynamics_2016,allen_contractile_2000,jiang_cellular_2013,venkova_mechano-osmotic_2021,rollin2023}. The evolutionary challenge to better regulate and control an organism's osmotic environment has led to a wide variety of different methods of osmoregulation. There is also much interest in the application and control of osmotic forces in \textit{in vitro} scenarios \cite{rangamani_lipid_2013}.

Many single celled organisms living in fresh water (such as bacteria and protists) are required to maintain a controlled volume and internal cytoplasmic concentration in order to function correctly. This is a challenge. The concentration of ions and proteins in their cytoplasm generates a much higher osmotic pressure than found in the surrounding environment, driving water  through the membrane and causing the cells to swell. This swelling leads to both an increase in the cell volume and a decrease in the ionic/protein concentration in the cytosol.

There are two main methods that can be used to combat this passive influx of water. The first is to build a rigid cell wall around the cell such that it can withstand the high pressure (of order the atmospheric pressure $\sim 100$kPa) generated by the imbalance of ionic concentrations across the cell membrane; this approach is typical of plant cells and most types of bacteria, where gram-positive typically use a thick cell wall and gram-negative use a thinner peptide-glycan network surrounding the cell \cite{wood_bacterial_2011,pilizota_origins_2014,amir_getting_2014}. This mechanism can be used to withstand pressures several times atmospheric pressure. Disruption of the synthesis machinery of these cell walls is the main target of many antibiotics.

A second option is to build an active pump which can remove excess water from the cell. This mechanism is typically favoured by fresh-water protists and amoeba that often have only an outer plasma membrane and no cell wall. The organelle responsible for this is called the contractile vacuole complex (CVC). This is a large, membrane-bound compartment lined with unidirectional proton or ion pumps which act to increase the ionic concentration inside the vacuole relative to the cytoplasm
\cite{allen_contractile_2000,patterson_contractile_1980}. The vacuole then inflates due to osmotic pressure imbalance between it and the cytoplasm. An opening in the plasma membrane then allows it to expel its contents into the extracellular medium. In this way it can transport water from the cytosol to the external environment, and thus give the cell a mechanism of osmoregulation. 

In this paper we will take inspiration from the structure of the contractile vacuole and ask how one might design a minimal cellular water pump. Utilising a simplified model of membrane tubes embedded with ion pumps we will show how a tube pulled into the cell interior with an open end connected, via the plasma membrane, to the exterior, can be used to pump water out of the cell under physiologically realistic pumping parameters. Beyond a critical pumping rate the tube undergoes an instability and a central vacuole forms, whose size is roughly that of the system. Finally we discuss how the structure could be used to osmoregulate and the requirements on the control of pumping rate needed to facilitate this. Throughout we take the approach of using a simple theory to give qualitative insight into these simple osmoregulation methods, rather than trying to predict precise quantitative details and morphology.

\section{The contractile vacuole complex}
The contractile vacuole complex (CVC) is a membrane-bound compartment found in many single-celled eukaryotes, in particular protists \cite{docampo_new_2013,allen_contractile_2000,patterson_contractile_1980}. The exact morphology of the CVC varies from species to species, here we will draw insipration from a common morphology exemplified by \textit{Paramecium} \cite{patterson_contractile_1980}. Here the CVC is formed of a main vesicle, connected to the plasma membrane by a large protein complex which forms a pore capable of opening and closing (most likely due to mechanical signalling). The main vesicle is covered in proton pumps which hydrolyse adenosine triphosphate (ATP) in order to move a proton across the membrane \cite{fok_pegs_1995}. This pumping causes an increased concentration of protons inside the main vesicle, which in turn allows for the flow of ions into the vesicle by passive channels (this highly simplified description of the electrophysiology does not accurately describe all details of the systems, \textit{e.g.}~cotransporters \cite{martinoia_vacuolar_2018,tominaga_electrophysiology_1998}, however it will suffice for our purposes). The overall increase in osmotic pressure (relative to the cytosol) generates a passive influx of water through aquaporin homologues \cite{stock_osmoregulation_2002}. The vesicle inflates until it reaches some critical point when the main pore opens and the fluid is expelled into the external medium.

In {\em Paramecium}, there are an array of membrane tubules connecting to a main vesicle, that inflate in a similar manner to the main vesicles  \cite{allen_contractile_2000}. These main tubes inflate and appear to undergo a peristaltic instability late in the inflation stage where they form bulges, called ampulla in the literature, of similar size to the main vacuole. This instability is driven by the osmotic imbalance from the ion pumps, it is qualitatively similar to a pearling instability driven by surface tension but has a natural wavelength several orders of magnitude larger. In previous work we studied a simple model for this instability, which leads to a robust long wavelength instability with wavelength set by the ratio of ion pumping time-scale to viscous time-scale, and the length-scale predicted by the theory is of the similar order of magnitude as the bulges seen in experiments on the CVC \cite{al-izzi_hydro-osmotic_2018}. It may be that these tubes act to more efficiently collect water, due to their large surface-to-volume ratio, which is transferred to the main vesicle as the tubes empty.

In this paper we will study a simplified model of a membrane tube pulled into the interior of a cell with an open end connecting the tube to the external medium. We will first study the fluid and ionic transport processes of such a tube with a fixed geometry at steady state, which will allow for a simple setting with which to understand the basic physics of the underlying transport processes. We show that for physiologically reasonable pumping rates it is possible for this steady state to expel water from the cell interior. Next we coupled this picture with a geometrically simplified model of membrane mechanics to allow for membrane deformation. Similar approaches have been used to great success in studying the physics and coarsening dynamics in lumen formation, however their potential as active water pumps has been largely ignored \cite{dasgupta_physics_2018,dumortier_hydraulic_2019,torres-sanchez_tissue_2021,verge_physics_2020}. We show that such a system not only forms a vacuole, due to the osmotic pressure imbalance, but it can also expel water from the cell. Finally we discuss possible feedback mechanisms by which this simple model could regulate cell volume.

\begin{figure}
\center\includegraphics[width=0.46\textwidth]{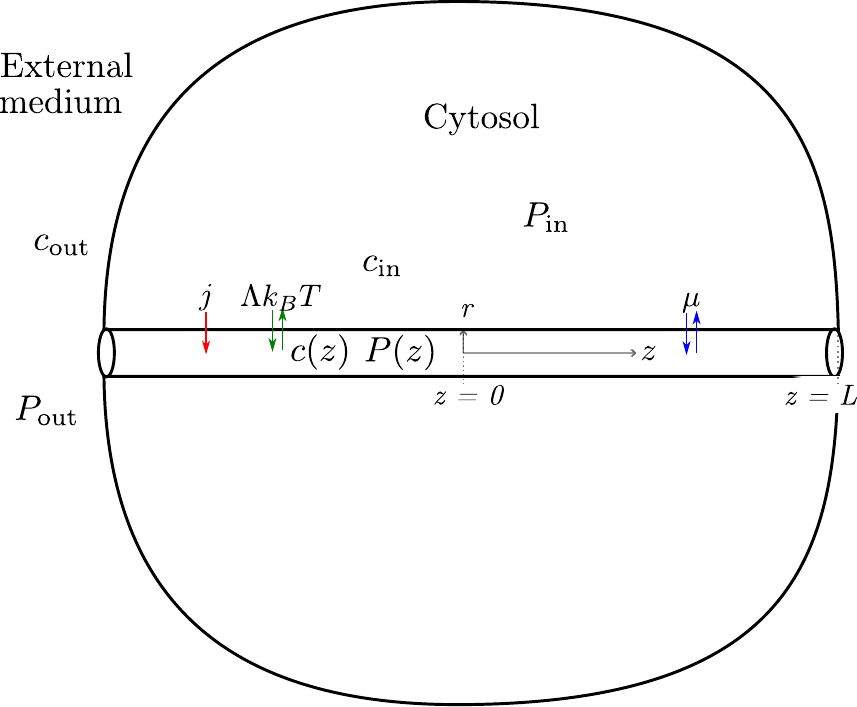}
\caption{\label{fig:schematic}Schematic showing a simple scenario in which a membrane tube (of constant radius) is pulled into the interior of a cell at hydrostatic pressure, $P_{\mathrm{in}}$, and has a concentration, $c_{\mathrm{in}}$, of osmolytes. The hydrostatic pressure and osmolyte concentration of the external medium is given by $P_{\mathrm{out}}$ and $c_{\mathrm{out}}$. The hydrodynamic pressure and osmolyte concentration inside the tube are given as a function of coordinate ($z$) along the length of the tube from the centre by $P(z)$ and $c(z)$ respectively. We assume that the tube has symmetric boundary conditions at the centre ($z=0$) for simplicity. Along the tube there are ion pumps which actively transport ions into the tube at a rate $j$ (red arrow), passive channels which allow ion flux driven by a difference in chemical potential via the coefficient $\Lambda k_{\text{B}}T$ (green arrows) and passive water channels which allow water to flow due to the pressure jump across the membrane with permeability $\mu$ (blue arrows). All transport parameters are per unit area. We assume uniform ionic concentration and hydrostatic pressure in the cytosol and external medium.}
\end{figure}

\section{Open-ended hydro-osmotic membrane tube}
First we consider an axisymmetric semi-permeable membrane tube of constant radius, $r$, and length, $L$, in the $z$ direction pulled into the interior of a cell at hydrostatic pressure, $P_{\mathrm{in}}$, and osmolyte concentration, $c_{\mathrm{in}}$, see Fig.~\ref{fig:schematic} for a schematic of this setup. We will assume neutral charge osmolytes throughout this paper, which for compactness we shall refer to as ions \cite{footnote2}. The purpose of this simple setup is to first illustrate some of the physical processes in a fixed geometry before moving on to more complex scenarios. This fixed geometry setup will allow us to more clearly examine some of the active processes, even if it is not mechanically feasible with real lipid bilayers. At $z=L$ we assume the tube is connected to the external medium at pressure, $P_{\mathrm{out}}$, and ionic concentration, $c_{\mathrm{out}}$. For simplicity we assume a reflective (Neumann) boundary condition at $z=0$, this simplifies some of the expressions, but does not change the qualitative nature of our results. More complex boundary conditions could be considered, for example, a tube with one open end held at fixed length or by a constant force.

The surface of the membrane tube is covered in uni-directional ion pumps pointed inward, and passive channels which can allow the flow of ions in either direction across the membrane. We will work in the steady state regime and balance the fluxes of ions and water at each cross section of the tube. The tube will be assumed to have a homogeneous water permeability per unit area, $\mu$, passive permeability to ions per unit area $\Lambda k_{\text{B}}T$ (where $k_{\text{B}}$ is Boltzmann's constant and $T$ is temperature) and an active pumping rate of ions per unit area $j$. Throughout this paper we assume neutral charge ions and neglect the electrostatics and pH, the later of which has been recently shown to have an effect on membrane tube morphology of possible biological relevance in mitochondria \cite{patil_mitochondrial_2020}.

The water flow down the tube is generated by gradients in the hydrodynamic pressure field, $P(z)$, the velocity profile, $\vec{u}\left(\vec{r}\right)$, at position $\vec{r}$ inside the tube is know as Poiseuille flow \cite{dommersnes_marangoni_2005} and is given by
\begin{equation}
u(\tilde{r}) = \frac{\partial_zP}{4\eta}\left(\tilde{r}^2-r^2\right);\quad Q(r) = -\frac{\pi}{8}\frac{r^4\partial_zP}{\eta}\mathrm{,}
\end{equation}
where $Q(r)$ is the flux integrated across the cross-section of the tube at a point $z$, $\eta$ is the viscosity and $\tilde{r}\leq r$ is the radial coordinate inside a tube of radius $r$.
This form can be found by the solution to the Stokes equations in cylindrical geometry with no-slip boundary conditions at the edge of the tube. If the fluid is incompressible then, $\partial_{z}Q+J_{w}=0$, with the flux of water through the boundary per unit length due to the passive water channels with permeation coefficient $\mu$ given by $J_{w}=2\pi r \mu \left[P-P_{\text{in}}-k_{\text{B}}T \left(c(z)-c_{\text{in}}\right)\right]$ where $P_{\mathrm{in}}$ and $c_{\mathrm{in}}$ are the pressure and ionic concentration inside the cell,
%and 
$c(z)$ is the ionic concentration in the tube, and ideal gas law is assumed for the ion osmotic pressure \cite{footnote}. From this we find the volume conservation equation
\begin{equation}
\lambda_{\mu}^2\partial_z^2P=P- k_{\text{B}}T c-(P_{\text{in}}-k_{\text{B}}T c_{\text{in}})\mathrm{,}
\label{pressure_eq}
\end{equation}
where $\lambda_{\mu}^2=r^3/16\eta\mu$. $\lambda_\mu$ is a length-scale describing the crossover between relaxation of hydrostatic pressure gradients by advective transport of water down the tube and by permeative transport of water across the membrane. We estimate $\lambda_\mu\simeq 100\mu$m, i.e. long compared with the size of the cell (see Table \ref{tab:lengthscales}).

We can write all the concentrations as osmotic pressure, as such $\Pi(z)=k_{\text{B}}T c(z)$. Flow down the tube is given by diffusion, with diffusion constant $D$, and advection with the fluid flow. Flow of ions across the membrane is given by two terms; the first is active pumping, $J_{i}^{(a)}=2\pi r j$, and the second is by passive leakage driven by the chemical potential difference across the membrane, which, assuming ideal gas law gives $J_{i}^{(p)}=2\pi r k_{\text{B}} T\Lambda \log\left[c/c_{\mathrm{in}}\right]$. If we balance diffusive and advective fluxes with the flow through the boundaries we can write an approximate ion conservation equation which at linear order about the ground-state $\Pi(z)=k_{\text{B}}T c_{\mathrm{in}}+\delta \Pi(z)$, $P(z)=P_{\mathrm{in}}+\delta P(z)$ gives
\begin{equation}
\lambda_D^2\partial_z^2 \delta\Pi + \lambda_{p}^2\partial_z^2 \delta P + \tilde{J} - \delta\Pi=0\mathrm{,}
\end{equation}
where $\tilde{J}= jc_{\text{in}}/\Lambda$, $\lambda_D^2 = Drc_{\text{in}}/2\Lambda k_{\text{B}} T$ and $\lambda_p^2=r^3c_{\text{in}}^2/16\eta\Lambda$. Note that the linear term in $\delta\Pi$ comes from expanding the logarithm term in the chemical potential. $\tilde{J}$ describes the ratio between active pumping and passive leakage of ions, and thus defines some equilibrium concentration in a uniform state in $z$ (\textit{e.g.}~a tube with a closed end). $\lambda_D$ describes the crossover between diffusive transport of ions at short length-scales and passive leakage at long length-scales, $\lambda_p$ describes a similar crossover for advective transport of ions vs passive ion leakage. Thus, the ratio $\lambda_p/\lambda_D$ is is a P\'{e}clet number for the system.

We can also rewrite pressure by expanding Eq.~\ref{pressure_eq} about the ground-state $P=P_{\text{in}}+\delta P$ giving
\begin{equation}
\lambda_{\mu}^2\partial_z^2\delta P=\delta P- \delta \Pi\mathrm{.}
\end{equation}

If we normalize the lengths by $\lambda_{\mu}$ and normalize pressure/osmotic pressure by $\Pi_{\mathrm{in}}=k_{\text{B}}T c_{\mathrm{in}}$ then we get the following dimensionless equations
\begin{align}\label{eq:tubeEqs1}
&\partial_z^2\delta P =\delta P -\delta \Pi\\ \label{eq:tubeEqs2}
&\xi_D^2 \partial_z^2\delta\Pi +\xi_p^2\partial_z^2 \delta P +J -\delta\Pi=0\text{,}
\end{align}
where $\xi_D=\lambda_D/\lambda_\mu$, $\xi_p=\lambda_p/\lambda_\mu$ and $J=j/\Lambda k_{\text{B}}T$. Estimates for these dimensionless parameters are given in Table \ref{tab:dimensionlessParams}. The underlying physical values from which these are estimated are given in Appendix \ref{sec:app1}.

\begin{table}[h!]
\begin{center}
\begin{tabular}{|c|c|}
\hline
Parameter & Value\\
\hline
$\xi_{D}=\sqrt{8D\eta\mu c_{\text{in}}/(k_{\text{B}}T\Lambda r^2)}$ & $0.5$\\
$\xi_p=c_{\text{in}}\sqrt{\mu/\Lambda}$ & $2.5$\\
$J=j/\Lambda k_{\text{B}}T$ & $[10^{-4}-10]$\\
\hline
\end{tabular}
\caption{\label{tab:dimensionlessParams}Table of values for dimensionless variables. Here $\xi_D$ is the dimensionless length-scale over which ions are screened by diffusion down the tube, $\xi_p$ is the length-scale over which ions are screened by advection and $J$ is the dimensionless rate at which ions are actively pumped into the tube.}
\end{center}
\end{table}

If we solve with the boundary conditions $\delta P(L)=P_{\mathrm{out}}-P_{\mathrm{in}}$, $\delta \Pi(L)=\Pi_{\mathrm{out}}-1$, $\partial_z\delta P|_{z=0}=\partial_z\delta\Pi|_{z=0} = 0$, then we find the following solution
\begin{widetext}
\begin{align}
\delta P(z) = &J + \cosh\left(\omega_-z\right)\frac{\left(P_{\text{out}}-P_{\text{in}} + 1 - \Pi_{\text{out}} + \left(J + P_{\text{in}} -P_{\text{out}}\right) \omega_+^2\right)}{\left(\omega_-^2-\omega_+^2\right)\cosh\left(\omega_-L\right)}\nonumber\\ 
&-\cosh\left(\omega_+z\right)\frac{\left(P_{\text{out}}-P_{\text{in}} + 1 - \Pi_{\text{out}} + \left(J + P_{\text{in}} -P_{\text{out}}\right) \omega_-^2\right)}{\left(\omega_-^2-\omega_+^2\right)\cosh\left(\omega_+L\right)}\mathrm{,}\\
\delta\Pi(z) = &J+ \left(1 - \omega_-^2\right) \cosh\left(\omega_-z\right)\frac{\left(P_{\text{out}}-P_{\text{in}} + 1 - \Pi_{\text{out}} + \left(J + P_{\text{in}} -P_{\text{out}}\right) \omega_+^2\right)}{\left(\omega_-^2-\omega_+^2\right)\cosh\left(\omega_-L\right)}\nonumber\\ 
&-\left(\omega_+^2-1\right)\cosh\left(\omega_+z\right)\frac{ \left(P_{\text{out}}-P_{\text{in}} + 1 - \Pi_{\text{out}} + \left(J + P_{\text{in}} -P_{\text{out}}\right) \omega_-^2\right)}{\left(\omega_-^2-\omega_+^2\right)\cosh\left(\omega_+L\right)}\mathrm{,}
\end{align}
\end{widetext}
where 
\begin{align}
\omega_{\pm}^2&= \frac{1}{2\xi_D^2}\left[1+\xi_D^2+\xi_p^2 \pm \sqrt{\left(1+\xi_D^2+\xi_p^2\right)^2-4\xi_D^2}\right]
%\mathrm{.}
\end{align}
With the numerical values given in Table \ref{tab:dimensionlessParams}., we find characteristic dimensionless inverse lengthscales of $\omega_+\simeq100$ and $\omega_-\simeq0.1$.

\begin{figure}
\center\includegraphics[width=0.48\textwidth]{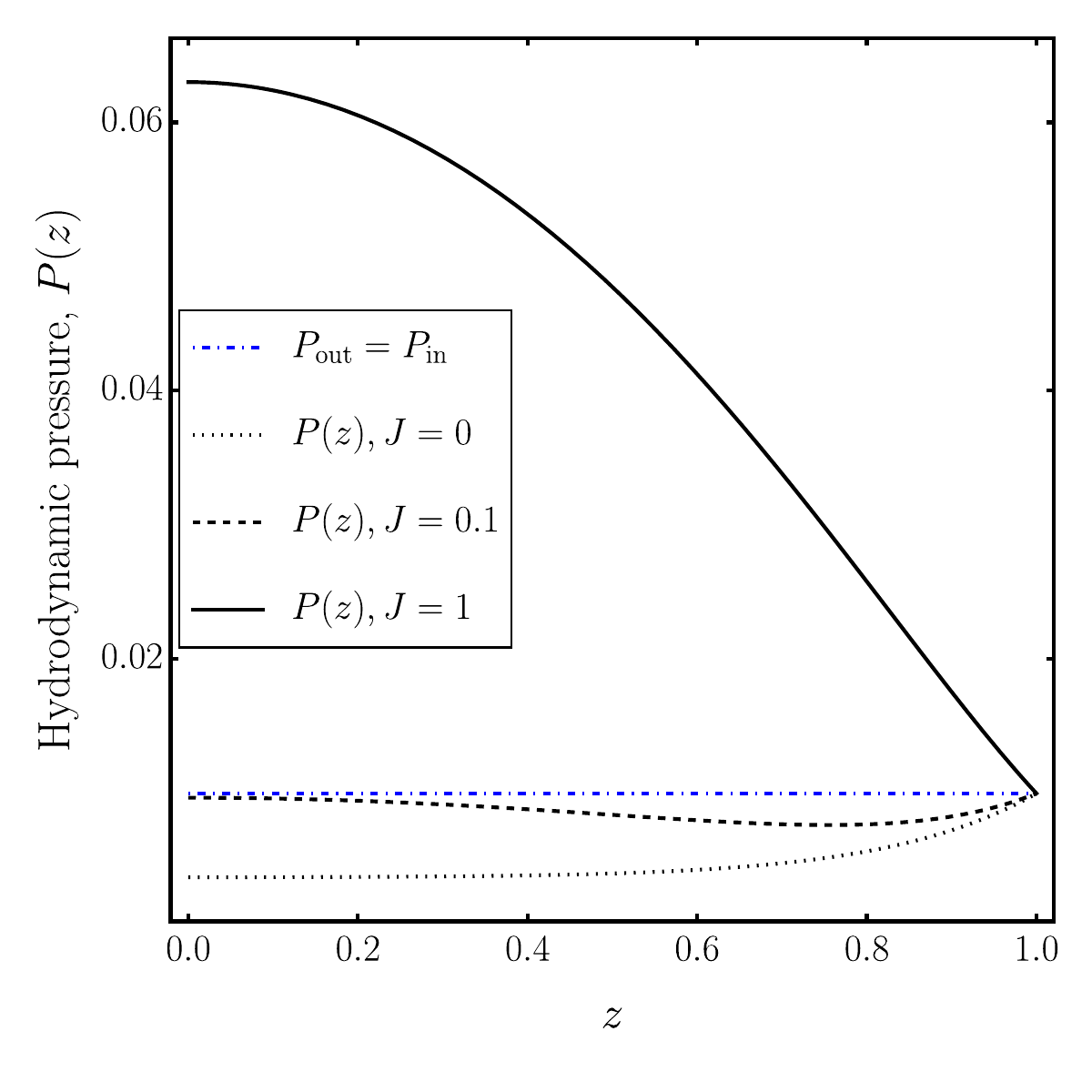}
\caption{\label{fig:constRadiusPressure}The pressure profile, $P(z)$, down a fixed geometry membrane tube with active ion pumps and passive channels as a function of position in the tube, $z$. 
The profiles are shown for several dimensionless pumping rates, $J$. The internal cell pressure/external medium pressure are $P_{\mathrm{in}}=P_{\mathrm{out}}=10^{-2}$ - blue dot-dashed line in (a). The pressures are in unit $\Pi_{\mathrm{in}}=k_{\text{B}}Tc_{\text{in}}$ and length scales in unit $\lambda_\mu$. Other parameters are set as $L=1$, $\xi_D=0.5$ and $\xi_{p}=2.5$. }
\end{figure}

\begin{figure}
\center\includegraphics[width=0.48\textwidth]{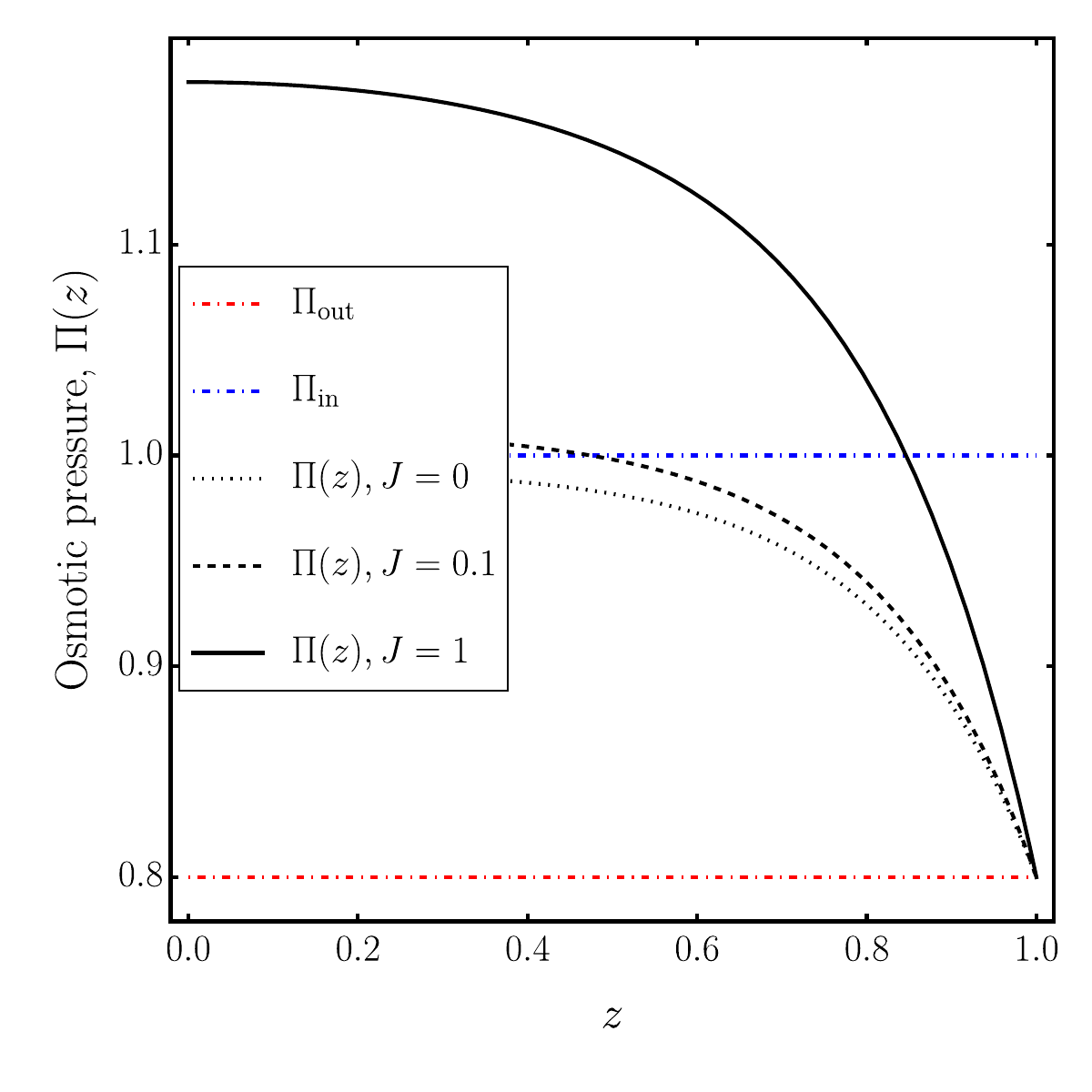}
\caption{\label{fig:constRadiusOsmoticPressure}The osmotic pressure profile, $\Pi(z)$, down a fixed geometry membrane tube with active ion pumps and passive channels as a function of position in the tube, $z$. 
The profiles are shown for several dimensionless pumping rates, $J$. The internal cell pressure/external medium pressure are $P_{\mathrm{in}}=P_{\mathrm{out}}=10^{-2}$ - blue dot-dashed line in (a). The internal cell osmotic pressure is $\Pi_{\mathrm{in}}=1$ (blue dot-dashed line) and the external osmotic pressure is $\Pi_{\mathrm{out}}=0.8$ (red dot-dashed line). The pressures are in unit $\Pi_{\mathrm{in}}=k_{\text{B}}Tc_{\text{in}}$ and length scales in unit $\lambda_\mu$. Other parameters are set as $L=1$, $\xi_D=0.5$ and $\xi_{p}=2.5$. }
\end{figure}

The solution for both pressures is plotted in Fig.~\ref{fig:constRadiusPressure} \&~\ref{fig:constRadiusOsmoticPressure}  for varying values of $J$. We choose the tube length to be $L=1$ in our dimensionless units, this corresponds to $L\sim 100\mu\text{m}$ in physical units (roughly the size of \emph{Paramecium} \cite{allen_contractile_2000}). For low pumping rate Fig.~\ref{fig:constRadiusPressure} shows a pressure profile that  drops on entering the tube from the external medium. This is due to water flowing into the tube through the open end and then permeating into the cell. This effect is screened on a length-scale of roughly $\xi_D$, determined by the balance between the diffusion of ions down the tube and the passive transport across the membrane. The ionic concentration in the tube, Fig.~\ref{fig:constRadiusOsmoticPressure}, is screened on this same length-scale beyond which the concentration is given by approximately the cytosolic ion concentration. At higher pumping rates the pressure profile increase in the centre of the tube scaling quadratically with position from the centre, Fig.~\ref{fig:constRadiusPressure}. This increase in hydrodynamic pressure is driven by a small increase in ionic concentration above the internal cell concentration along the length of the tube, the cumulative effect of which leads to a large increase of pressure at the center of the tube, see Fig.~\ref{fig:constRadiusPressure} \& \ref{fig:constRadiusOsmoticPressure}.

The flux of water out the end of the tubes is given, in dimensionless units, by 
\begin{align}
&\tilde{Q}|_{z=L}= \frac{Q}{2\pi r \mu k_{\text{B}} T \lambda_\mu^3 c^{\text{in}}} = \frac{\xi_p^2}{\omega_+^2 - \omega_-^2}\times\nonumber\\
&\Big\{(J+\Delta P)\omega_+\omega_-(\omega_+ \tanh (L \omega_-)-\omega_- \tanh (L \omega_+))\nonumber\\
&-(\Delta\Pi-\Delta P)(\omega_+ \tanh (L \omega_+)-\omega_- \tanh (L \omega_-))\Big\}\mathrm{,}
\end{align}
where $\Delta\Pi=1-\Pi_{\text{out}}$ and $\Delta P=P_{\text{in}}-P_{\text{out}}$. In order for the tube to be able to expel water the following condition must be satisfied 
\begin{equation}
\tilde{Q}|_{z=L}\propto-\partial_zP|_{z=L}>0\text{.}
\end{equation}

\begin{figure}
\center\includegraphics[width=0.48\textwidth]{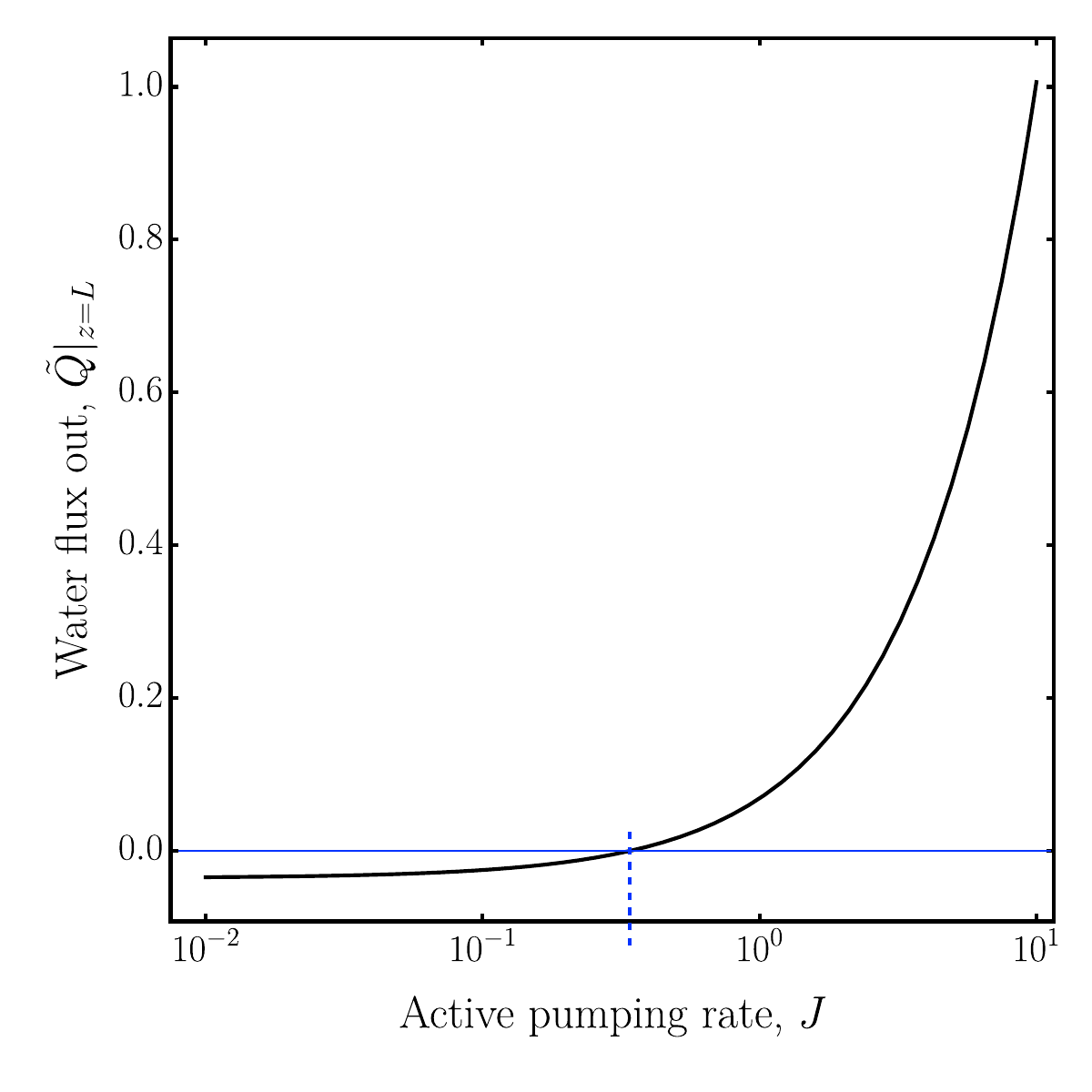}
\caption{\label{fig:waterFlux} The flux of water out the open end of the tube, $\tilde{Q}|_{z=L}$, as a function of pumping rate, $J$, for internal osmotic pressure $\Pi_{\mathrm{in}}=1$, external osmotic pressure $\Pi_{\mathrm{out}}=0.8$, internal and external hydrostatic pressure $P_{\mathrm{in}}=P_{\mathrm{out}}=10^{-2}$ (all in dimensionless units, see main text for details). The critical flux allowing to expel water from the cell (Eq.\ref{jcrit}) is shown (dashed line). Other parameters are set as $L=1$, $\xi_D=0.5$ and $\xi_{p}=2.5$.}
\end{figure}

We plot this flux as a function of $J$ in Fig.~\ref{fig:waterFlux}. This shows a monotonic increase in water flux with pumping rate from an initially negative value (water flowing into the cell via the tube) to a point where the flux becomes positive and tube starts to expel water. It is interesting to note that this flux also depends linearly on the difference in osmotic pressure $\Delta\Pi=1-\Pi_{\mathrm{out}}$ (with a negative coefficient).

\begin{figure}
\center\includegraphics[width=0.48\textwidth]{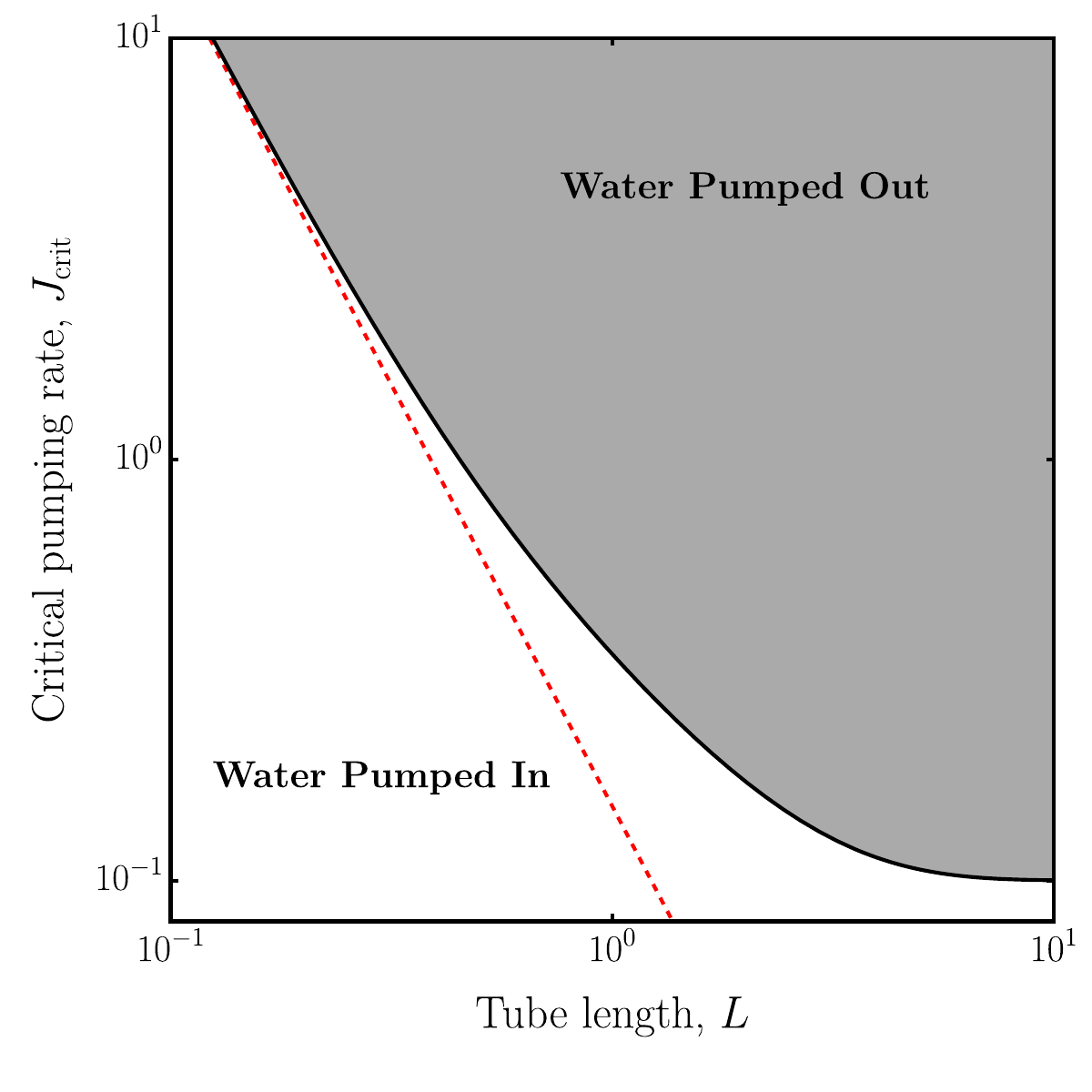}
\caption{\label{fig:phaseDiagram} Phase diagram of influx/out-flux for the critical pumping rate for the flux to be out of the tube, $J_{\mathrm{crit}}$, against length of tube. The dashed red line indicates the criterion in the limit that passive ionic permeation, $\Lambda$, is small. Parameters given by $\Pi_{\mathrm{in}}=1$, external osmotic pressure $\Pi_{\mathrm{out}}=0.8$, internal and external hydrostatic pressure $P_{\mathrm{in}}=P_{\mathrm{out}}=10^{-2}$ (all in dimensionless units, see main text for details). Other parameters are set as $\xi_D=0.5$ and $\xi_{p}=2.5$.}
\end{figure}

The critical pumping rate to expel water is found when the pressure gradient at $z=L$ changes sign. This is given by
\begin{align}
J_{\mathrm{crit}}=&\frac{(\Delta\Pi-\Delta P)(\omega_+ \tanh (L \omega_+)-\omega_- \tanh (L \omega_-))}{\omega_+\omega_-(\omega_+ \tanh (L \omega_-)-\omega_- \tanh (L \omega_+))}\nonumber\\
&-\Delta P\text{,}
\label{jcrit}
\end{align}
which is plotted as a function of tube length, $L$ in Fig.~\ref{fig:phaseDiagram}. 
In the physiologically relevant case where $\omega_-\ll1$ and $\omega_+\gg1$, we find $J_{\mathrm{crit}}\sim L^{-1}$ for $\omega_-L\,\ll1\ll\omega_+L$, see the dashed red line in Fig.~\ref{fig:phaseDiagram} For large tube length and keeping passive permeation fixed the critical pumping rate becomes
\begin{equation}
J_{\mathrm{crit}}\approx\frac{1-\Pi_{\text{out}}-(1+\omega_-\omega_+)(P_\text{in}-P_\text{out})}{\omega_-\omega_+}\mathrm{.}
\end{equation}

This corresponds to the asymptote of $J_{\mathrm{crit}} \approx \frac{\Pi_\text{in}-\Pi_{\text{out}}}{\omega_-\omega_+}\approx 0.1$ in Fig.~\ref{fig:phaseDiagram}. This finite value of pumping needed to expel water in the infinite tube limit is due to the screening of the osmotic pressure by the passive ion channels. Note that for values of lengths similar to the size of \emph{Paramecium}, $L\sim 1$, we are in a regime where passive permeation cannot be neglected.

Throughout this we have assumed a tube of constant radius, which is an assumption that breaks down as soon as a small pressure difference exists across the tube's membrane as the radius will deform and the tube will undergo a pearling-like instability, as discussed in \cite{al-izzi_hydro-osmotic_2018}. This instability is similar to a Rayleigh-Plateau instability in a column of viscous fluid \cite{rayleigh_xvi._1892,tomotika_instability_1935}, but typically has a much longer wavelength set by the ratio of the pumping and viscous timescales \cite{al-izzi_hydro-osmotic_2018}. We discuss the corrections due to a small change in the tube radius in Appendix \ref{sec:app2} and show that the simple linear approximation breaks down very fast, thus in the next section we turn to a simple toy model to account for the geometric non-linearities.

\section{Hydro-osmotic membrane tube with a central vacuole}
\begin{figure}
\center\includegraphics[width=0.46\textwidth]{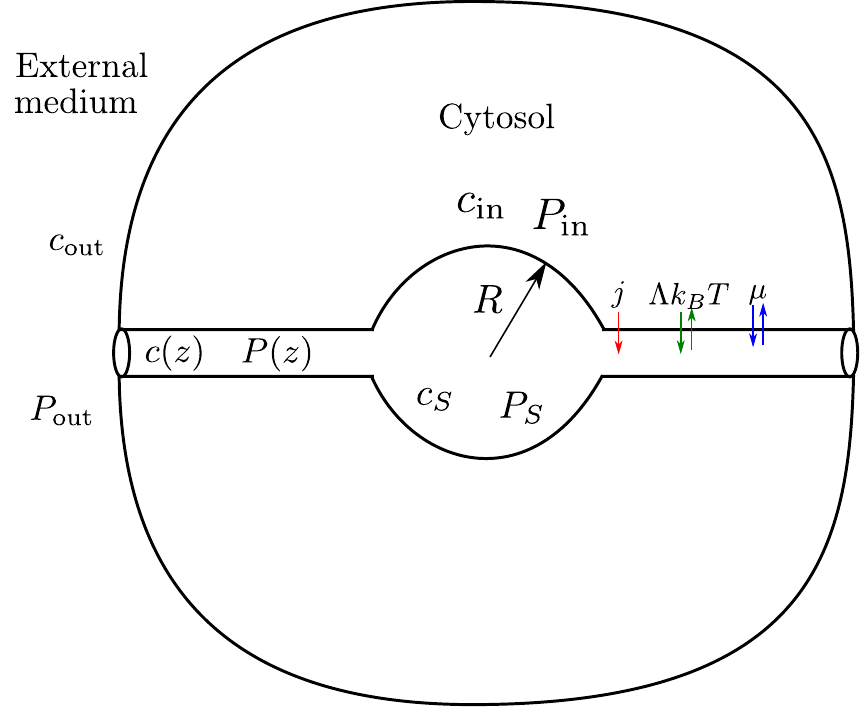}
\caption{\label{fig:schematic2}Schematic showing the second system we consider where the setup is the same as Fig.~\ref{fig:schematic}, but now we allow for the formation of a vacuole at the centre of radius, $R$, pressure $P_{S}=P_\text{in}+\frac{2\sigma}{R}$ (with $\sigma$ the surface tension of the vacuole) and concentration $c_{S}$, which are assumed constant within the vacuole. All other parameters are as in Fig.~\ref{fig:schematic}.}
\end{figure}

We can now consider a simplified tube+sphere model where instead of the Neumann boundary conditions at the centre of the tube we have $\delta P(z=R)=P_{S}-P_{\text{in}}$, $\delta\Pi(z=R)=\Pi_{S}-\Pi_{\text{in}}$ where $P_{S}$, $\Pi_{S}$ are the pressure and osmotic pressure in a sphere of radius $R$ centered at $z=0$ (assumed to be uniform in the sphere), see Fig.~\ref{fig:schematic2} for a schematic of this setup. The pressure in the sphere is given by $P_{S}=P_{\text{in}}+2\sigma/R$ where $\sigma$ is the membrane surface tension \cite{safran_statistical_1994}. This setup is highly simplifed and neglects the energy of the neck joining the tube and the sphere, however such approaches have proved a useful way to analyse similar \textit{in vitro} scenarios \cite{dommersnes_marangoni_2005}. We show how this crude approximation is in qualitative agreement with the full shape equation in the limit of large but slowly varying radius, and also that the force balance in the tube yields a radius that is independent of pressure at lowest order in the derivatives in radius, see Appendix \ref{sec:app3}. We solve Eq.~\ref{eq:tubeEqs1},~\ref{eq:tubeEqs2} for the pressure and osmotic pressure in a tube with these boundary conditions which yields the solutions found in Appendix \ref{sec:app4}.

We now balance the fluxes in the sphere to find the radius $R$ and internal osmotic pressure $\Pi_S$. Thus we obtain the following flux balance equation for water flux in dimensionless form (with dimensionless tension $\tilde{\sigma}=\sigma/\Pi_{\text{in}}\lambda_\mu$)
\begin{equation}\label{eq:tubeVacuoleVolume}
R^2\left(\Pi_S-1-\frac{2\tilde\sigma}{R}\right)+ r \partial_zP|_{z=R}=0\text{,}
\end{equation}
and for the ion flux
\begin{equation}\label{eq:tubeVacuoleIon}
R^2\left(J-\Pi_S+1\right) + \xi_p^2 r\partial_zP|_{z=R}+ \xi_D^2 r \partial_z\Pi|_{z=R}=0\text{.}
\end{equation}
The first term in Eqs.\ref{eq:tubeVacuoleVolume} \& \ref{eq:tubeVacuoleIon} are the fluxes from the cytosol into the vacuole across the vacuole membrane, and the second term(s) are the fluxes between the vacuole and the tube.

\begin{figure}
\center\includegraphics[width=0.48\textwidth]{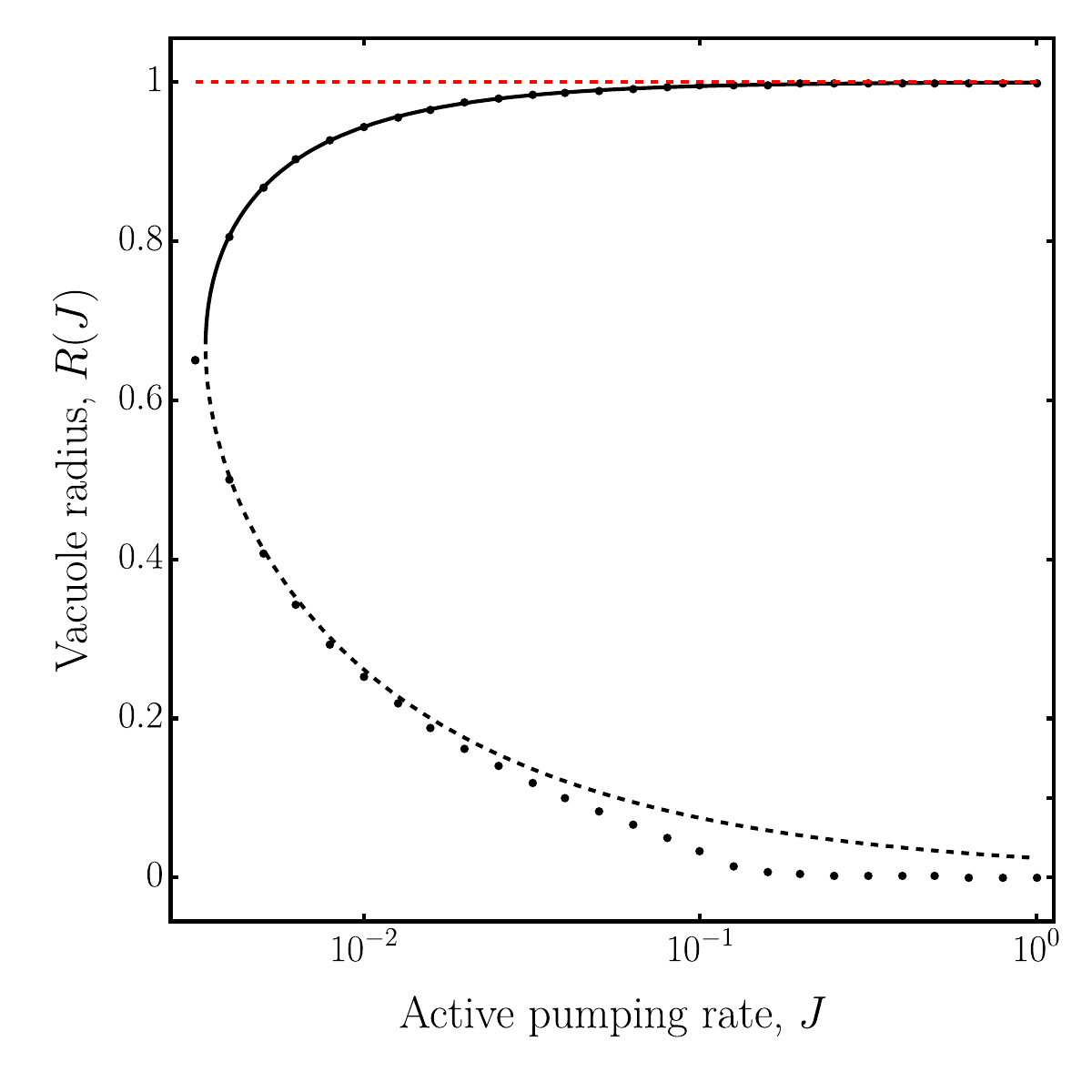}
\caption{\label{fig:vacuole}Stable (solid) and unstable (dashed) radii, $R$, for a spherical vacuole at the centre of the tube shown in Fig.~\ref{fig:schematic2} plotted against dimensionless ion pumping rate, $J$. 
The dots represent the full numerical solution and the lines the approximate solution for large vacuole. The dashed red line indicates the cell edge. Other parameters are set as $L=1$, $r=0.01$, $\xi_D=0.5$ and $\xi_{p}=2.5$.}
\end{figure}

The osmotic and hydrostatic pressure profiles inside the tubes and the water outflux through the tube end can be obtained analytically for a given vacuole radius $R$, see Appendix \ref{sec:app4}. 
We then find the roots to Eqs.~\ref{eq:tubeVacuoleVolume} \& \ref{eq:tubeVacuoleIon}   numerically for $R$ and $\Pi_s$ to find the steady state, Fig~\ref{fig:vacuole}. To simplify the problem we can find an approximate solution to the steady state in the limit where the vacuole is close to the edge giving the following approximate fluxes
\begin{align}\label{eq:waterFluxApprox}
&\partial_zP|_{z=R}\approx\frac{P_{\text{out}}-2\tilde\sigma/R-P_{\text{in}}}{L-R}\mathrm{,}\\
&\partial_z\Pi|_{z=R}\approx\frac{\Pi_{\text{out}}-\Pi_{S}}{L-R}\mathrm{.}\label{eq:ionFluxApprox}
\end{align}

This can then be substituted into Eqs.~\ref{eq:tubeVacuoleVolume},~\ref{eq:tubeVacuoleIon} where the roots to the equations can be found using a symbolic programming language, such as Mathematica (Wolfram Research, Champaign, IL). This approximation is in qualitative agreement with the full solution even when the vacuole is not close to the edge, Fig~\ref{fig:vacuole}.
%dots (full solution) vs lines (approximate solution).

The real roots for the radius, $R$, are plotted as a function of pumping rate, $J$, Fig.~\ref{fig:vacuole} showing a saddle-node bifurcation at $J\sim 10^{-2.5}$ for this choice of parameters. This can be understood from the tube elastic energy becoming unstable as it undergoes an inflection point for large pressure. This gives a stable branch which corresponds to a vacuole \blue{}that rapidly fills the system size as the pumping rate is increased and an unstable branch that is not physically realisable. Making use of the approximate fluxes (Eq.~\ref{eq:waterFluxApprox} \& \ref{eq:ionFluxApprox}) and the fact that, for our choice of parameters, $\xi_D/\xi_p$ is small can give a lowest order approximation for the vacuole radius as $R^3 (L-R) \approx \frac{2\tilde{\sigma} r_0}{J}(1+\xi_p^2)$ which tends to the system size as $J\to \infty$.

Note that we are using a constant pumping flux and permeabilites per unit area here, corresponding to a constant density of pumps and channels independent of the vacuole size. An alternative approach is to consider an ensemble where the total number of pumps is fixed. In this case, the pumping rate needed to form a vacuole is substantially higher, but the qualitative features of the system remains the same and the two ensembles converge in the limit of high pumping rate.

\begin{figure}
\center\includegraphics[width=0.48\textwidth]{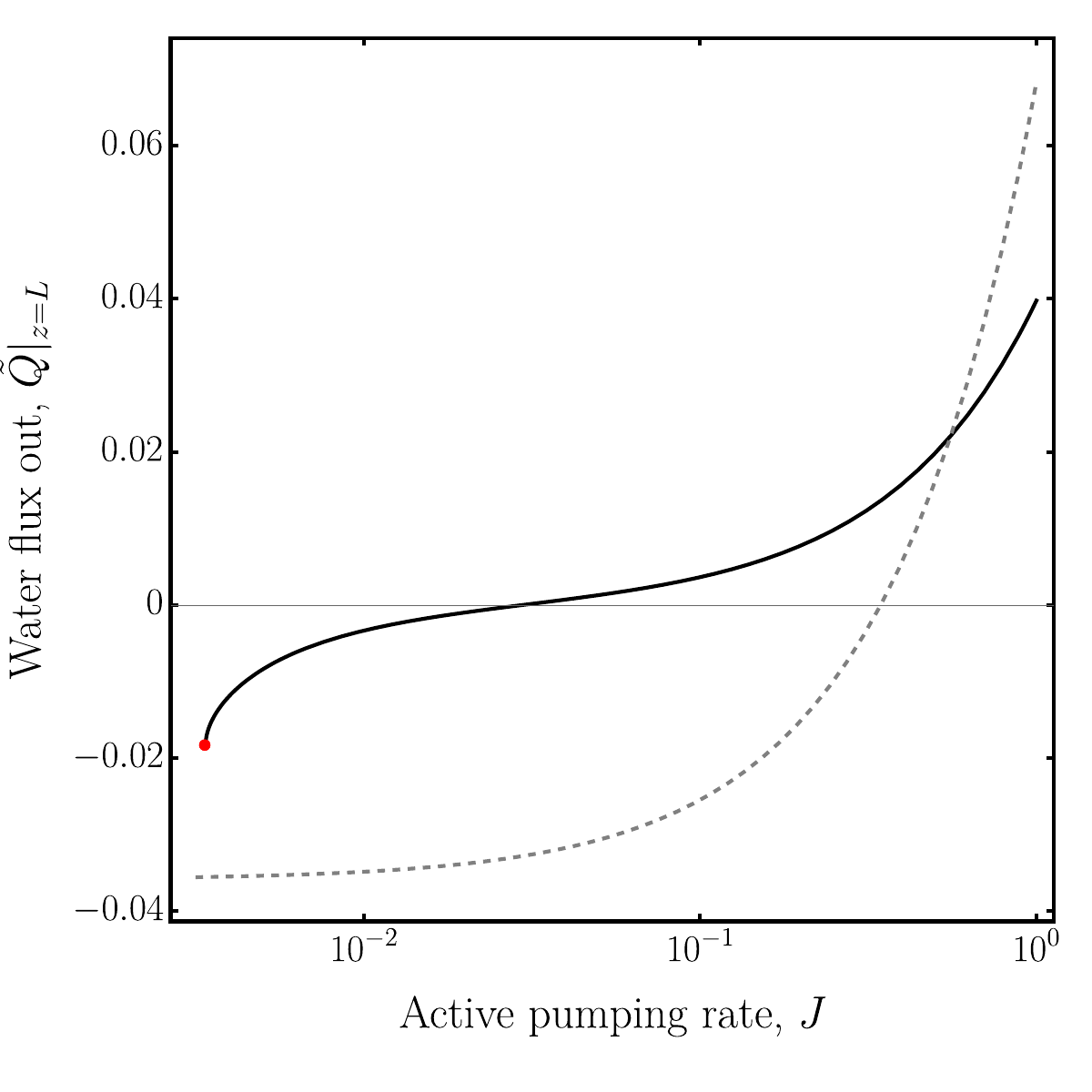}
\caption{\label{fig:vacuoleWaterFlux} The flux of water out the open end of the tube, $\tilde{Q}|_{z=L}$, as a function of pumping rate, $J$, for internal osmotic pressure $\Pi_{\mathrm{in}}=1$, external osmotic pressure $\Pi_{\mathrm{out}}=0.8$, internal and external hydrostatic pressure $P_{\mathrm{in}}=P_{\mathrm{out}}=10^{-2}$ (all in dimensionless units, see main text for details). Other parameters are set as $L=1$, $r=0.01$, $\xi_D=0.5$ and $\xi_{p}=2.5$. The dashed line shows the case of a fixed tube geometry for comparison and the red dot shows the bifurcation where the vacuole is formed.}
\end{figure}

The flux of water out of the end of the tube can be plotted as a function of pumping rate, $J$, see Fig.~\ref{fig:vacuoleWaterFlux}. After the vacuole is formed there is a jump in the total active pumping due to the large increase of surface area of the vacuole. This increased ion flux enables the vacuole to pump out water for a much lower active pumping rate when compared to the fixed geometry active membrane tube (more that an order of magnitude with our parameter estimates, see Fig.~\ref{fig:vacuoleWaterFlux}). Interestingly this benefit to the vacuole formation for the active water pump does not hold at large values of active pumping rate. As $J$ becomes large the vacuole size is limited by the size of the system and eventually the fixed tube shape would be able to pump more water out due to the much higher pressures generated at the centre of the tube, Fig.~\ref{fig:vacuoleWaterFlux}. Although conceptually interesting, this situation is physically inplausible.

\section{Discussion}
In this paper we have shown how a simple toy model of a membrane tube covered in unidirectional active ion pumps and bidirectional passive water and ion channels can form a cellular water pump able to expel water from the cytosol into the external medium. The osmotic pressure imbalance naturally leads to the formation of a vacuole highly reminiscent of the contractile vacuole complex, a organelle responsible for osmoregulation in many protists \cite{allen_contractile_2000,more:2024}. This vacuole formation has the benefit of lowering the active pumping criterion needed in order to expel fluid. This suggests that a membrane tube, or network of membrane tubes might provide a way of generating such a vacuole \textit{de novo} simply through the mechanics of osmotic pressure imbalance.

Crucially, this outflux of water provides a minimal way in which a single cell can regulate its volume. Note that, when comparing the active outflux to the passive influx of water from the extracellular medium ($J^{\mathrm{water}}_{\mathrm{in}}\sim A_{\mathrm{cell}}\Delta\Pi$) we can see how for sufficiently high pumping it should be possible for the active water flux out to balance the passive osmotic flux into the cell. Of course, in order for the cell to actively respond to changes in osmotic pressure an active feedback mechanism is needed \cite{jiang_cellular_2013}.

A possible example of this, that has some precedent in the plant science literature, would be for the pumping rate to depend on cellular calcium concentration \cite{lew_calcium_1989}, as the low levels of calcium within cells make them highly sensitive to small increases in its concentration. It is well know that many freshwater protists contain mechano-sensitive calcium channels that allow an influx of calcium ions when a critical surface tension of the plasma membrane is reached, such channels were also recently show to be important for collective hydrodynamic signalling in \textit{Spirostomum} \cite{mathijssen_collective_2019}. 
Furthermore, hypo-osmotic or calcium-rich external conditions trigger extra contractile vacuole complex generation in \emph{Paramecium multimicronucleatum} \cite{iwamoto:2003}.
If an influx of calcium could lead to an increased activity of the proton pumps (as discussed in other context in \cite{yao:2003}), then it is possible that this could lead to a minimal method of osmoregulation. The details of this feedback mechanism and any fine tuning required are beyond the scope of the current paper and we leave for future work.

Throughout this work we have restricted ourselves to minimal models to elucidate the basic physics of these membrane tube based water pumps. A possible extension would be to solve the full mechanical and diffusive non-linear PDEs to better characterize the morphology of such vacuoles by using finite element approaches \cite{arroyo2009,sahu2020,barrett2016,reuther2020,torres-sanchez2019}. 

We have also considered purely steady state solutions so a further interesting extension, which dovetails with the finite element approach might be to simulate the full hydrodynamics of the system and look for regimes with transient pulsatile behaviour similar to that seen in Ref.~\cite{dasgupta_physics_2018} in the case of lumen formation in cell-cell aggregates/tissues.

Finally, and perhaps of more interest biologically, would be to integrate more precise detailed electro-physiological models \cite{rollin2023} to understand the role of electrostatics for osmoregulation in living freshwater organisms with no cell wall and the role of electro-osmotic effects on membrane morphology.

\appendix

\section{Estimates for parameters}\label{sec:app1}

We first consider  the  known physical parameters in our model, we detail their values in Table \ref{tab:basicParams}. We assume diffusion of ions in water, viscosity on a similar order of magnitude to water, and typical cellular pressures and ion concentrations along with a typical size for a membrane tube.
\begin{table}[h!]
\begin{center}
\begin{tabular}{|c|c|c|}
\hline
Parameter & Value & Ref.\\
\hline
$D$ & $10^3\mu\text{m}^2\cdot\text{s}^{-1}=10^{-9}\mathrm{m}^2\cdot\text{s}^{-1}$ & \cite{milo_cell_2015}\\
$\eta$ & $10^{-3}\mathrm{Pa}\cdot\text{s}$ & \cite{milo_cell_2015}\\
$c_{\text{in}}$ & $100\text{mM}$ ($c_{\text{in}}k_{\text{B}}T\sim 10^{5}\mathrm{Pa}$) & \cite{milo_cell_2015}\\
$r$ & $0.1\mu\text{m}$ & \cite{derenyi_formation_2002,zhong-can_bending_1989}\\
$P_{\text{in}}$ & $10^3\text{Pa}\sim P_{\text{out}}$ & \cite{milo_cell_2015}\\
$k_{\text{B}} T$ & $10^{-21}\mathrm{J}$ & \cite{milo_cell_2015}\\
\hline
\end{tabular}
\caption{\label{tab:basicParams}Table of basic physical parameter values.}
\end{center}
\end{table}

Assuming that the ion channels are fast ($\sim 10^7$ ions passing through one channel a second) and ion pumps are slow ($\sim 10^4$ ions per pump per second) and that they have a density of order $1$ every $10 \mathrm{nm}^2$ we can estimate the transport coefficients for the ions. For the membrane permeability to water we can find values in the literature. These values are given in Table \ref{tab:transportCoefficitents}.
\begin{table}[h!]
\begin{center}
\begin{tabular}{|c|c|c|}
\hline
Parameter & Value (order of magnitude) & Ref.\\
\hline
$j$ & $10^6-10^9 \mathrm{\mu m^{-2}s^{-1}}=10^{18}-10^{21}\mathrm{m^{-2}.s^{-1}}$ & \cite{stock_osmoregulation_2002,al-izzi_hydro-osmotic_2018}\\
$\mu$ & $10^{-13}-10^{-10}\mathrm{m}\cdot\text{Pa}^{-1}\cdot\text{s}^{-1}$ & \cite{olbrich_water_2000}\\
$\Lambda$ & $10^{44}-10^{45}\mathrm{J}^{-1}\cdot\text{m}^{-2}\cdot\text{s}^{-1}$ & \cite{milo_cell_2015}\\
\hline
\end{tabular}
\caption{\label{tab:transportCoefficitents}Table of values for transport/active pumping coefficients.}
\end{center}
\end{table}

Finally we include a table summarizing the three natural lengthscales of the system, $\lambda_\mu$, $\lambda_D$ and $\lambda_p$ in Table~\ref{tab:lengthscales}.

\begin{table}[h!]
\begin{center}
\begin{tabular}{|c|c|}
\hline
Parameter & Value (order of magnitude)\\
\hline
$\lambda_{\mu}$ & $100\mu\text{m}$\\
$\lambda_{D}$ & $50\mu\text{m}$\\
$\lambda_p$ & $250\mu\text{m}$\\
\hline
\end{tabular}
\caption{\label{tab:lengthscales}Table of values for natural length-scales and pumping/leakage ratio}
\end{center}
\end{table}

These estimates give dimensionless parameters as follows $\xi_{D}\sim10^{-1}$, $\xi_p\sim 10$ and $J\sim 10^{-6}-10^{-1}$. However, we note that these estimates made from basic quantities only give a crude approximation to the complex electrophysiological processes occuring in real cells. The reality is that many of the active pumps seen in real vacuolar systems are proton pumps which then act to contransport other ions into the vacuole \cite{martinoia_vacuolar_2018}. If we treat our dimensionless parameters as coarse grained parameters approximating the complex underlying electrophysiology then we can in fact generate much higher pumping rates. If we take ionic concentrations from electrophysiology on real contractile vacuole complexes \cite{tominaga_electrophysiology_1998} and consider a simple tube now with a closed end (as the mechano-sensitive pore in real contractile vacuoles acts to close the structure to the external medium \cite{allen_contractile_2000}) then we find a steady-state where the active ion pumping is given by $J=k_{\text{B}} T\Lambda \delta c/c_{\mathrm{in}}$. This leads to much higher estimates for $J\sim 10^{-1}-1$. In the main paper we consider the full range of these values. 

\section{Tube with linearised variable radius}\label{sec:app2}
The set of dimensionless ODEs for a tube with small variable radius $r(z)=r_0\left(1+\delta r\right)$ and surface tension $\sigma=\sigma_0\left(1+\delta \sigma\right)$ are given by
\begin{align}
&P_r\left(2\delta r +2\xi_0^4\partial_z^4\delta r +\delta \sigma\right)=\delta P\text{,}\\
&\partial_z^2\delta P=\delta P-\delta\Pi \mathrm{,}\\
& \xi_D^2\partial_z^2 \delta\Pi + \xi_{p}^2\partial_z^2 \delta P + J\left(1+\delta r\right) - \delta\Pi=0\text{,}\\
&2P_r\partial_z\delta\sigma=\partial_z\delta P\text{,}
\end{align}
where $P_r=\sigma_0/r_0c_{\mathrm{in}}k_\mathrm{B}T$ and $\xi_0=r_0/\lambda_{\mu}$. Above, Eq.B1 is the normal force balance at the tube membrane, Eqs.B2-B3 are the balances of flux for ions and water, and Eq.B4 is the tangential force balance between the membrane tension gradient and the fluid shear stress. We choose $\sigma_0=\kappa/2 r_0^2$ where $\kappa$ is the membranes bending rigidity. As such the membrane is in force balance for $\delta P=0$. By noting that $\xi_0\ll 1$ for our parameters we can make the approximation
\begin{equation}
P_r\left(2\delta r +\delta \sigma\right)\approx\delta P\text{,}
\end{equation}
and we can integrate the tangential force balance equation to find
\begin{equation}
\delta \sigma = \frac{1}{2P_r}\left(\delta P+ P_{\mathrm{in}}-P_{\mathrm{out}}\right)\mathrm{.}
\end{equation}

Our approximate solution for the radius is thus given by
\begin{equation}
\delta r = \frac{1}{4P_r}\left(\delta P +P_{\mathrm{out}}-P_{\mathrm{in}}\right)\mathrm{,}
\end{equation}
as $P_{r}\sim 10^{-5}$ this implies that very small changes in pressure can lead to a large deformation of the tube. Because of this the linear regime of the shape equation is likely to be inadequate to describe the mechanics of the deformable tube. Rather than solve the full non-linear equations numerically we appeal to a crude approximation of a fixed radius tube connected with a sphere, this is justified by the sphere being a high pressure limit of the shape equation for a tube with slowly varying (but not small) radius, see Appendix \ref{sec:app3}. Such approaches have also been shown to work in similar \textit{in vitro} systems \cite{dommersnes_marangoni_2005}.

\section{Arguments for the simplified tube-vacuole model}\label{sec:app3}
\subsection{Spherical vacuole as the high pressure limit of a weakly non-linear tube}
Here we consider a tube with slowly varying radius and show that under high pressure it reduces to a spherical geometry. Consider a spatially varying radius of the form
\begin{align}
&r(z)=r_0 S\left(z\right) \\
& \partial_z^n S\sim \mathcal{O}(\epsilon^n) \quad \mathrm{where} \quad \epsilon\ll 1\text{.}
\end{align}

First we consider the shape equation, expanding in powers of $\epsilon$. The full shape equation is given by
\begin{align}\label{eq:fullShape}
&\kappa \left[2 \Delta_{\mathrm{LB}} H -4 H\left(H^2-K\right)\right] + 2 \sigma H\nonumber\\
& +\Delta P=0\text{,}
\end{align}
where $\Delta_{\mathrm{LB}}$ is the Laplace-Beltrami operator, $\kappa$ is the bending rigidity and $\sigma$ the surface tension this gives a characteristic radius $r_0=\sqrt{\kappa/(2\sigma)}$ and $\Delta P$ is the pressure jump accross the membrane.

Up to second order in $\epsilon$ we find
\begin{align}
&H\approx\frac{-1}{2r_0 S}+ \frac{r_0}{2}\left[S_{zz}+\frac{S_{z}^2}{2S}\right]\mathrm{,}\\
&K\approx \frac{-S_{zz}}{S}\mathrm{,}\\
&2\Delta_{LB}H \approx \frac{S_{z}^2}{r_0 S^3} -\frac{S_{zz}}{r_0 S^2}\mathrm{.}
\end{align}

Substituting these expressions into Eq.~\ref{eq:fullShape} gives a slowly varying shape equation of the form
\begin{align}
&\frac{1}{2S^3}-\frac{\tilde{\sigma}}{S}+\xi_0^2\left\{\frac{S_{z}^2}{4S^3} -\frac{1}{2}\frac{S_{zz}}{S^2}\right\}\nonumber\\
& + \xi_0^2 \tilde{\sigma}\left[S_{zz}+\frac{S_{z}^2}{2S}\right]+ \frac{1}{P_\kappa}\left(P-P_{\text{in}}\right)=0\text{,}
\end{align}
where $\xi_0=r_0/\lambda_{\mu}$ (and $\lambda_{\mu}=\sqrt{r_0^3/(16\eta\mu)}$) as we have renormalised all our length-scales by $\lambda_{\mu}$. The dimensionless surface tension is given by $\tilde{\sigma}=\sigma r_0^2/\kappa$ (note that the surface tension is in general not constant but will depend on external traction forces on the membrane), $P_{\kappa}=\kappa/r_0^3c_{\text{in}}k_{\text{B}}T$ and pressure is non-dimensionalised in terms of the interior osmotic pressure of the cell $c_{\text{in}}k_{\text{B}}T$. Noting that $\xi_0$ is small we can make the observation that for $S\gg 1$ the first two terms dominate the equation. From large $S$ all the terms multiplied by $\xi_0$ except one scale like $S^{-n}$ (for $n\geq 1$). We can thus reduce the equation to
\begin{equation}
\frac{1}{2S^3}-\frac{\tilde{\sigma}}{S} + \xi_0^2 \tilde{\sigma}S_{zz} + \frac{1}{P_\kappa}\left(P-P_{\text{in}}\right)=0\text{.}
\end{equation}
For very large pressure the radius increases and the first term can be neglected, the equation is now solved by $S(z)=\sqrt{R^2-(z/\xi_0)^2}$ (assuming terms that go like $S^{-3}$ can be neglected) and a surface tension given by the usual Laplace pressure of a spherical vesicle in the central region of the tube. This is simply the equation for a spherical cap centred at $z=0$, thus our crude approximation seems to match with this more formal expansion of the full shape equation, at least close to the centre of the tube ($z=0$).

\subsection{Constant tube radius check}
As an additional check that our assumption of a constant tube radius that is independent of pressure once the vacuole is formed is reasonable we consider a simplified model of the energetics of a tube and sphere under pressure. The free energy of the system is given by
\begin{align}
\mathcal{F}=&-\Delta P_S \left(\pi  r^2 (L-R)+\frac{4 \pi  R^3}{3}\right)\nonumber\\
&+\frac{\pi \kappa (L-R)}{r}+2 \pi r \sigma (L-R) + 4 \pi R^2 \sigma
\end{align}
where the length of the tube is given by $L-R$ and $\Delta P_S=P_S-P_{\text{in}}$, is the pressure jump across the vacuole membrane, with $P_S=P_\text{in}+2\sigma/R$, as in the main text. If we assume the sphere is large, $r\ll R$, then a simple minima to this energy is given by
\begin{equation}
r=\sqrt{\frac{\kappa}{2\sigma}};\quad R= \frac{2\sigma }{\Delta P_S}\text{,}
\end{equation}
which shows that the tube radius, $r$, is independent of the hydrostatic pressure, thus justifying our constant tube radius approximation.

\section{Solutions for tube-vacuole model}\label{sec:app4}
Here we record the analytical solutions to the pressure and osmotic pressure inside a tube of length $L-R$ in the combined tube with a spherical vacuole problem. The boundary conditions at $z=R$ are given by $\delta P(z=R)=P_{S}-P_{\text{in}}$, $\delta\Pi(z=R)=\Pi_{S}-\Pi_{\text{in}}$ where $P_{S}$, $\Pi_{S}$ are the pressure and osmotic pressure in a sphere of radius $R$. The pressure in the sphere is given by $P_{S}=P_{\text{in}}+2\sigma/R$ where $\sigma$ is the membrane surface tension. The full solutions are the following,
\begin{widetext}
\begin{align}
&\delta P(z) = \frac{1}{2 R (\omega_{-}^2-\omega_{+}^2)}\Bigg[\mathrm{csch}(\omega_{-} (L-R)) \mathrm{csch}(\omega_{+}(L-R)) \bigg\{2 \sinh (\omega_{-} (L-R))\nonumber\\
&\times \big[R \sinh (\omega_+ (R-z)) \left(\omega_-^2 (J+P_{\mathrm{in}}-P_{\mathrm{out}})-P_{\mathrm{in}}+\Pi_{\mathrm{in}} + P_{\mathrm{out}} -\Pi_{\mathrm{out}}\right) \nonumber\\
&-\sinh (\omega_+ (L-z)) \{R \left(J \omega_-^2+ \Pi_{\mathrm{in}} - \Pi_{S}\right)-2 \sigma  \omega_-^2 + 2 \sigma \}\big]\nonumber\\
& +2 \omega_+^2 \sinh (\omega_+ (L-R)) ((J R-2 \sigma ) \sinh (\omega_- (L-z))-R (J+P_{\mathrm{in}}-P_{\mathrm{out}}) \sinh (\omega_- (R-z)))\nonumber\\
&+J R \left(\omega_+^2-\omega_-^2\right) \cosh ((L-R) (\omega_- -\omega_+))+J R (\omega_-^2 -\omega_+^2) \cosh ((L-R) (\omega_- + \omega_+ ))\bigg\}\nonumber\\
&+\frac{4 e^{\omega_- (L+R)}}{e^{2 L \omega}-e^{2 R \omega_-}} \Big[\sinh (\omega_- (L-z)) (R (\Pi_{\mathrm{in}}-\Pi_{S})+2 \sigma )\nonumber\\
&+R (P_{\mathrm{in}}-\Pi_{\mathrm{in}}- P_{\mathrm{out}} + \Pi_{\mathrm{out}}) \sinh (\omega_- (R-z))\Big]\Bigg]\mathrm{,}
\end{align}
\begin{align}
&\delta\Pi(z) = J + \frac{1}{R (\omega_-^2-\omega_+^2)}\Bigg\{\left(\omega_-^2-1\right) \mathrm{csch}(\omega_- (L-R))\nonumber\\
&\times \Big[R \sinh (\omega_- (R-z)) \left(\omega_+^2 (J+P_{\mathrm{in}}-P_{\mathrm{out}}) - P_{\mathrm{in}} + \Pi_{\mathrm{in}} + P_{\mathrm{out}} - \Pi_{\mathrm{out}}\right)\nonumber\\
&-\sinh (\omega_- (L-z)) \left(\omega_+^2 (J R-2 \sigma )+R (\Pi_{\mathrm{in}} - \Pi_{S}) + 2 \sigma \right)\Big]+\left(\omega_+^2-1\right) \mathrm{csch}(\omega_+ (L-R))\nonumber\\
&\times\Big[\sinh (\omega_+ (L-z)) \left(\omega_-^2 (J R-2 \sigma )+R (\Pi_{\mathrm{in}} - \Pi_{S})+2 \sigma \right)\nonumber\\
&-R \sinh (\omega_+ (R-z)) \left( \omega_-^2 (J+ P_{\mathrm{in}} - P_{\mathrm{out}})- P_{\mathrm{in}} + \Pi_{\mathrm{in}} + P_{\mathrm{out}} -\Pi_{\mathrm{out}}\right)\Big]\Bigg\}\mathrm{,}
\end{align}
\begin{align}
&\tilde{Q}|_{z=L}=\frac{-\xi_p^2}{2 R (\omega_-^2-\omega_+^2)}\Bigg\{2 \omega_+ \Big[R \coth (\omega_+ (R-L)) \left(\omega_-^2 (J+ P_{\mathrm{in}}-P_{\mathrm{out}})- P_{\mathrm{in}}+ \Pi_{\mathrm{in}} + P_{\mathrm{out}} - \Pi_{\mathrm{out}}\right)\nonumber\\
& + \omega_- \omega_+ \mathrm{csch}(\omega_- (L-R)) (R (J+ P_{\mathrm{in}}- P_{\mathrm{out}}) \cosh (\omega_- (R-L))-J R+2 \sigma )\nonumber\\
&+\mathrm{csch}(\omega_+ (L-R)) \left(\omega_-^2 (J R-2 \sigma )+R (\Pi_{\mathrm{in}} - \Pi_{S})+2 \sigma \right)\Big]\nonumber\\
&+\frac{4 \omega_- e^{\omega_- (L+R)} (R \cosh (\omega_- (R-L)) (-P_{\mathrm{in}}+ \Pi_{\mathrm{in}}+ P_{\mathrm{out}}-\Pi_{\mathrm{out}})+R (\Pi_{S}-\Pi_{\mathrm{in}})-2 \sigma )}{e^{2 L \omega_-}-e^{2 R \omega_-}}\Bigg\}\mathrm{.}
\end{align}
\label{big_expressions}
\end{widetext}

\section*{Acknowledgements}
S.C.A.-I.~was partially supported by EPSRC under grant number EP/L015374/1, Labex CellnScale (ANR-11-LABX-0038) part of the IDEX PSL (ANR-10-IDEX-0001-02 PSL) and the EMBL-Australia program.

%\bibliographystyle{unsrt}

%\bibliography{bibliography}

\begin{thebibliography}{99}
\bibitem{footnote2} One could view the active pumping of passive osmolytes in this model as the active pumping of one species of charged ion along with the passive flow of counterions needed to maintain charge balance. We leave the full solution of this electro-hydro.osmotic problem to future work.
\bibitem{footnote} If we assume that the tube is much narrower than it is long and that osmolyte diffusion is fast on length-scales of the tube radius then we can write, $c(\vec{r})=c(z)$.
\end{thebibliography}

\end{document}